\documentstyle[amsfonts,pre,aps,epsfig,float,delarray]{revtex}

\draft

\def\uni{u_n^{i}}
\def\un1i{u_{n+1}^{i}}
\def\vni{v_n^{i}}
\def\vn1i{v_{n+1}^{i}}

\def\vecr{{\mathbf r}}
\def\vecm{{\mathbf F}}
\def\manm{\mathcal{M}}
\def\vecz{{\mathbf z}}
\def\vecd{{\mathbf e}}
\def\vecf{{\mathbf G}}
\def\jacobi{{\mathbf J}}
\def\vece{{\mathbf e}}
\def\tni{\cot\theta^{i}_n}
\def\transpose{^{\sf T}}

\def\figw{0.48\textwidth}

\begin{document}

\wideabs{

\begin{flushright}
\sf SOGANG-ND 79/01
\end{flushright}

\title{Mechanism of Synchronization in a Random Dynamical System}

\author{Dong-Uk Hwang,$^{1,2,a}$ Inbo Kim,$^{1,2,b}$  Sunghwan Rim,$^{1,c}$
    Chil-Min Kim,$^{1,d}$ and Young-Jai Park$^{2,e}$}

\address{$^1$National Creative Research Initiative Center for Controlling
  Optical Chaos,\\
  Department of Physics, Paichai University, Seogu, Daejeon
  302-735, Korea}
\address{$^2$Department of Physics and Basic Science Research Institute,
    \\ Sogang University, Seoul 100-611, Korea}

\date{\today}

\maketitle

\begin{abstract}
The mechanism of synchronization in the random Zaslavsky map is
investigated. From the error dynamics of two particles, the
structure of phase space was analyzed, and a transcritical
bifurcation between a saddle and a stable fixed point was found. We
have verified the structure of on-off intermittency in terms of a
biased random walk. Furthermore, for the generalized case of the
ensemble of particles, \emph{a modified definition} of the size of
a snapshot attractor was exploited to establish the link with a random walk.
As a result, the structure of on-off intermittency in the ensemble
of particles was explicitly revealed near the transition.

\end{abstract}


}

\narrowtext
%%%%%%%%%%%%%%%%%%%%%%%%%%%%%%%%%%%%%%%%%%%%%%%%%%%%%%%%%%%%%%%%%%%%%%%
%%%%%%%%%%%%%%%%%%%%%%%%%%%%%%%%%%%%%%%%%%%%%%%%%%%%%%%%%%%%%%%%%%%%%%%
\section{Introduction}
\label{sec:int}
%Introduction part

Yu, Ott, and Chen (YOC) studied a transition to chaos for a random
dynamical system with the random Zaslavsky (RZ) map which
describes the motion of particles floating on the surface of a
fluid whose flow velocity has complicated time
dependence\cite{yoc:prl,yoc:physica}; the state of the system is being
sampled at discrete times. It was shown that variation
of a parameter causes a transition from a situation, where an
initial cloud of particles is eventually distributed on a fractal,
to a situation, where the particles eventually clump at a single
point, whose location moves randomly in all time. In both situations
the random motion persists permanently, so the concept
of attractor is inappropriate. Hence, after transient times had elapsed, 
they took a snapshot of the particle distribution on the fluid
surface, and called it \textit{a snapshot
attractor}\cite{yoc:prl,romeiras:pra:90}. They observed that the
long-time particle distribution that evolves from an initial
smooth distribution exhibits an ``extreme form of temporally
intermittent bursting'' on the chaotic side near the transition.

After YOC's work, a type of intermittent behavior known as on-off
intermittency has been reported by Platt, Speigel and
Tresser\cite{platt:prl:93}. On-off intermittency refers to the
situation where some dynamical variables exhibit two distinct
states in their course of time evolution. One is the `off' state,
where the variable remain approximately a constant, and the other
is the `on' state, where the variable temporarily burst out of the
off state. It has long been thought that the intermittent behavior
in YOC's work belongs to ``on-off intermittency''
\cite{yang:pre:96,rim:prl:00,ding:pre:95,lai:pre:96-1,ding:pre:97,lai:pre:96-2,qu:pre:96,stefanski:pre:98,kye:pre:01}.
However, there has been some confusion about this because YOC did
not investigate the RZ map from the perspective of on-off
intermittency. Indeed in the Letter\cite{yoc:prl} and subsequent
review paper\cite{yoc:physica}, they analyzed the RZ map by
studying the largest Lyapunov exponent, the size of the snapshot
attractor, and the simple one-dimensional contraction-expansion
random map model to understand the intermittent transition
behavior. But the essential geometrical structure related to the
mechanism of on-off intermittency is absent in their analysis.
Besides Heagy, Platt, and Hammel (HPH) commented that the snapshot
attractor can undergo a form of intermittent behavior that is
similar to on-off intermittency, but the size distribution of
the snapshot attractor computed by YOC is quite different from the
laminar phase distribution they obtained in their random walk
model\cite{heagy:pre:94}.

Meanwhile, Yang and Ding (YD) have studied a
noise-driven uncoupled map lattice as a spatially extended
system\cite{yangding:pre:94,snapshot:other}. As a special case,
they considered a map lattice where logistic map is located at
each site uncoupled with its neighbors and each map driven by a
common random variable, which is regarded as a homogeneous
background. In their work, similar transition which comes from the
instability of the synchronous motion of the ensemble was found.
However, on-off intermittency of the size evolution of the snapshot
attractor was proposed merely on the ground of laminar distribution
obtained from numerical calculations since YD were unable to map the
size evolution of the snapshot attractor into a random walk. In this
respect, their study was incomplete.

Recently synchronization in a pair of nonlinear systems subjected
to the common noise is revisited\cite{rim:prl:00,fahy:prl:92,maritan:prl:94,gade:pla:96}.
Gade and Basu showed that this synchronization phenomena is indeed
physical in certain cases, and for synchronization of those system
the randomness is not vital\cite{gade:pla:96}. Moreover we have
shown that, in this kind of synchronization, the distribution of
the random variable is vital and there exists on-off intermittency in
the boundary of synchronization region\cite{rim:prl:00}. Also the
intermittent behavior of the size of snapshot attractor was
discussed briefly. Note that
synchronization of random dynamical systems  
has same meaning of the chaotic transition in YOC's work.
This notion of sychronization is analogous to the
synchronization scheme\footnote{ The scheme is based on the fact that
certain chaotic systems possess a self-synchronization property.  A
Chaotic system is self-synchronizing if it can be decomposed into 
subsystems: a drive subsystem and a stable response subsystem that
synchronize when coupled with a common drive signal.} 
proposed by Pecora and Carrol,
if we assume that random variable is realized by a 
system as a drive system and random dynamical systems are regarded as a
response systems\cite{pecora:prl:90}.

In this paper, we explicitly show that the size evolution
of the snapshot attractor really has the structure of on-off
intermittency contrary to the HPH's comment. In Section II, we
briefly recapitulate the properties of the RZ map, and in
Sec.~\ref{sec:sync} the mechanism of synchronization of two
particles will be revealed by using the error dynamics. In
Sec.~\ref{sec:randomwalk}, we introduce modified definition of the
size of the snapshot attractor for the case of ensemble of particles,
and explicitly show that this modification leads to the structure
of on-off intermittency. Finally we  present summary and brief
discussion in Sec.~\ref{sec:summary & discussion}.

%%%%%%%%%%%%%%%%%%%%%%%%%%%%%%%%%%%%%%%%%%%%%%%%%%%%%%%%%%%%%%%%%%%%%%%%%%%%%%
%%%%%%%%%%%%%%%%%%%%%%%%%%%%%%%%%%%%%%%%%%%%%%%%%%%%%%%%%%%%%%%%%%%%%%%%%%%%%%
\section{Properties of the Random Zaslavsky Map}
\label{sec:prop}

A generic $D$-dimensional random map can be described as follow:
\begin{eqnarray}
  \label{eq:general random map}
  \vecr_{n+1} = \vecm_{\xi_n} (\vecr_n),
\end{eqnarray}
where $\vecr_n\in{\Bbb R}^D$ is column state vector and $\xi_n$ is
a random variable. So the map $\vecm_{\xi_n}$ is chosen randomly
at each iteration according to some rule generating $\xi_n$. As it
was introduced in \cite{yoc:prl,yoc:physica}, the RZ map describes
a particle floating on an incompressible fluid. There is constant
divergence transverse to the surface, and this divergence leads
to contraction on the surface. And there exists vortical flow with
complicated time dependence. $\vecr_n=(x_n,y_n)\transpose$ (the
superscript `${\sf T}$' means transpose operation) describes the
position of a particle on the surface at $t=nT$, where $T$ is a
constant sampling time. $\vecm_{\xi_n}$ is given by
\begin{eqnarray}
  \label{eq:yoc map}
  \begin{array}[c]{l}
    x_{n+1}=x_n + f(\alpha)y_n {\rm~~mod~~}2\pi \\[2mm]
    y_{n+1}=g(\alpha)y_n+k\sin(x_{n+1}+\xi_n),
  \end{array}
\end{eqnarray}
where $f(\alpha)=(1-e^{-\alpha})/\alpha$, $g(\alpha)=e^{-\alpha}$,
and $k$ and $\alpha$ are control parameters. In specific, $\alpha$
gives rate of constant contraction of surface, $k$ stands for
the parameter of vortical flow, $x$ is angle variable, and $y$
is radial variable. On the other hand, random variable
$\xi_n$ gives complicated time dependence of vortical flow, which
represents fluid instabilities at low Reynolds numbers. When
$\xi_n$ is absent, Eq.~(\ref{eq:yoc map}) is often called the
Zaslavsky map\cite{zaslavsky:pla:78}. If we consider a strip
$|y|<K_0$ where $K_0>k(1-e^{-\alpha})^{-1}$, one iteration maps this
strip into a narrow band $|y|<K_1<K_0$. Thus long term behavior is
confined to $|y|<k(1-e^{-\alpha})^{-1}$. Note that Jacobian
determinent of the map is $g(\alpha)=e^{-\alpha}$. Therefore, the
map is contracting by the factor $g(\alpha)$ after each iteration.
In this paper, we will restrict our interest to the case of $k=0.5$ as
studied by YOC.
Given a uniform distribution of the ensemble of particles on the
surface (two-dimension), we can find fractal distribution of
particles for $\alpha<\alpha_c\simeq0.31$ after several iterations.
When we increase $\alpha$ close to $\alpha_c$, snapshots show
almost one-dimensional fractal structure (very thin line
structure). And for the case of $\alpha>\alpha_c$, the snapshot
collapses to a point (zero-dimension), and particles move
synchronously as if they were a single particle.

\section{Synchronization of two particles}
\label{sec:sync}

From now on, we will consider particles which
satisfy Eq.~(\ref{eq:yoc map}). Because $\xi_n$ is describing
geometry of vortical flow at $t=nT$ and particles are sprinkled on
the surface, they have different initial conditions but feel
common random driving $\xi_n$. To investigate the synchronization
of particles, we will begin to study the dynamical behavior of two
particles rather than the motion of whole ensemble of particles.

Let us consider a replica $\vecr'_n=(x'_n,y'_n)\transpose$ which
obeys the Eq.~(\ref{eq:yoc map}) as does $\vecr_n$. To show the
mechanism of synchronization, we consider the error dynamics (ED)
of this system, $\vecd_n=(u_n,v_n)\transpose=\vecr'_n-\vecr_n$,
which is the difference of two arbitrarily chosen particle's positions. Then,
the whole system $\vecz_n = (x_n,y_n,u_n,v_n)\transpose$  can be
described by
\begin{eqnarray}
  \label{eq:whole system}
  \begin{array}[c]{cl}
    \vecr_{n+1}&=\vecm_{\xi_n}(\vecr_n),\\[2mm]
    \vecd_{n+1}&=\vecf_{\phi_n}(\vecd_n),
  \end{array}
\end{eqnarray}
where $\vecf_{\phi_n}$ is given by
\begin{eqnarray}
    \label{eq:error-dynamics}
    \begin{array}[c]{l}
      u_{n+1}=u_n + f(\alpha)v_n ~\overline{\bmod}~ 2\pi   
      \\[2mm]
      v_{n+1}=g(\alpha)v_n +
      2k\cos\left(\phi_n+{u_{n+1}\over 2}\right)
      \sin\left({u_{n+1}\over 2}\right)
    \end{array}
\end{eqnarray}
with $[A ~\overline{\bmod}~ B]\equiv (A+B/2 \bmod B)-B/2$ and
$\phi_n\equiv x_{n+1}+\xi_n$. Here, we have applied mod operation
according to the translational symmetry of $u_n$ in trigonometric
functions in Eq.~(\ref{eq:error-dynamics}).
Note that the ED is driven by $\phi_n$, which is the sum of
$x_{n+1}$ from $\vecm_{\xi_n}(\vecr_n)$ and random variable
$\xi_n$. Therefore $\phi_n$ is not affected by the evolution of
$\vecd_n$, hence transformed system have so-called skew-product
structure mathematically\cite{platt:prl:93}. Because the
trajectory near hyperplane $\vecd=0$ is governed by $\phi_n$, we will study
properties of the ED regarding the random dynamical variable $\phi_n$ as a
system parameter $\phi$ which is constant during the evolution, before
considering whole four-dimensional system.

\subsection{Error Dynamics of two particles}
\label{subsec:ed}
\begin{figure}[htbp]
  \centering
  \epsfig{file=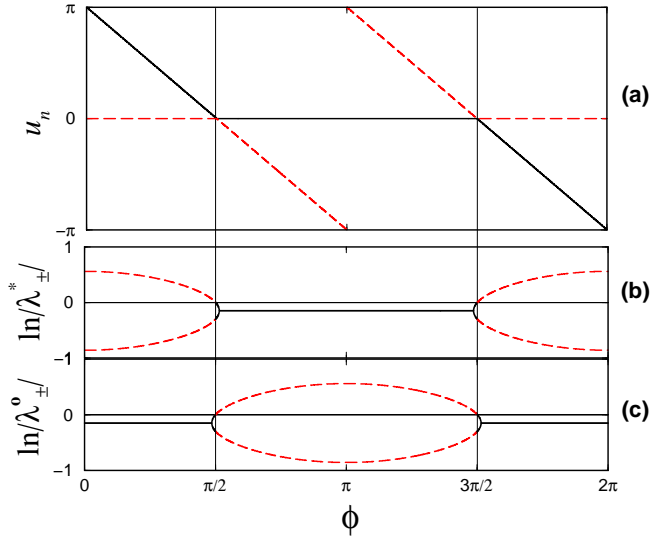,width=\figw}
  \caption{(a) Bifurcation diagram of the ED.
  In this figure $v_n$ is omitted because fixed
  point is always $0$. Horizontal axis is $\phi$ and its
  value is same as in (b) and (c). In the diagram, solid and dashed
  lines stand for stable and unstable fixed points respectively.
  (b) and (c) show logarithm of eigenvalues
  ($\lambda^*_\pm,\lambda^\circ_\pm$) of Jacobian
  matrix at $\vecd^*$ and $\vecd^\circ$, respectively. Vertical lines at
  $\phi={\pi/ 2},{3\pi/ 2}$ stand
  for transition points of fixed points where $\ln|\lambda^*_+|$ and
  $\ln|\lambda^\circ_+|$ become 0.}
  \label{fig:bifur}
\end{figure}

%%%%%%%%%%%%%%%%%%%%%%%%%%%%%%%%%%%%%%%%%%%%%%%%%%%%%%%%%%%%%%%
%
% properties of ED and bifurcation
%
We will investigate the local and global bifurcation structure by
finding the fixed points of $\vecf_\phi$ and its phase portrait in order to
reveal the underlying structure of the ED. When we impose condition
$u_{n+1}=u_n$ and $v_{n+1}=v_n$ on $\vecf_\phi$, $v_nf(\alpha)$
should be integer multiple of $2\pi$. If $v_n=0$,
\begin{eqnarray}
\label{eq:fixed point condition}
    \cos\left(\phi+{u_n \over 2} \right) \sin\left( {u_n \over 2} \right)=0.
\end{eqnarray}
Then, there are two fixed points,
\begin{eqnarray}
\label{eq:fixed points}
    \vecd^*&=& \left( \begin{array}{c} u^* \\ v^* \end{array} \right)
    = \left( \begin{array}{c} 0 \\ 0 \end{array} \right
    )~~\mbox{and}\nonumber\\[2mm]
    \vecd^\circ&=& \left( \begin{array}{c} u^\circ\\ v^\circ
    \end{array} \right) =
    \begin{array}\{{lll}.
      \left(\begin{array}{c} \pi-2\phi\\ 0\end{array}\right)
      & \mbox{for} & 0\leq\phi<\pi\\[5mm]
      \left(\begin{array}{c} 3\pi-2\phi\\ 0 \end{array}\right)
      & \mbox{for} & \pi\leq\phi<2\pi.
    \end{array}
\end{eqnarray}
When $v_nf(\alpha)$ equals
$\pm 2\pi,\pm 4\pi \cdots$, there is no additional fixed points
near transition point $\alpha_c$\cite{ed-fp}. The
$\vecd_{n+1}$ near the fixed point 
$\vecd^*$ is written as follows:
\begin{eqnarray}
  \label{eq:linear map}
  \vecd_{n+1}-\vecd^*
  = \jacobi^*(\phi) (\vecd_n-\vecd^*) + \ldots ,
\end{eqnarray}
where $\jacobi^*(\phi)=(\partial \vecf_\phi / \partial
\vecd)|_{\vecd^*}$ is Jacobian matrix evaluated at $\vecd^*$. Two
eigenvalues $\lambda^*_+$, $\lambda^*_-$ and the corresponding
eigenvectors $\vece_+$, $\vece_-$ can be obtained by solving
eigenvalue problem of the matrix $\jacobi^*(\phi)$. 
% Because an arbitrary
% $\vecd_0$ can be decomposed into $c_+ \vece_+$ and $c_- \vece_-$,
%\begin{eqnarray}
%  \label{eq:decomposed eq}
%  \vecd_n=(\lambda^*_+)^n c_+ \vece_+ + (\lambda^*_-)^n c_- \vece_-.
%\end{eqnarray}
%For the particular case of $\vecd_0=c\vece_\pm$, $\vecd_n$ will be
%attracted to $\vecd^*$
%when $|\lambda^*_\pm|<1$, while repelled when $|\lambda^*_\pm|>1$.
%Therefore 
Pairs of eigenvalues give behavior along eigendirection
of $\vecd_\pm$, and characterize the stability of $\vecd^*$
\cite{moon:book}. From Eq.~(\ref{eq:error-dynamics}), Jacobian
matrix is given by
\begin{eqnarray}
\label{eq:jacobian1}
      \jacobi^*(\phi)=\left(
        \begin{array}{cc}
        1 & f(\alpha) \\
        k\cos \phi ~ & ~ g(\alpha)+kf(\alpha)\cos \phi
        \end{array}\right).
\end{eqnarray}
And its eigenvalues are
\begin{eqnarray}
  \label{eq:eigenvalue1}
  \lambda^*_\pm &=& {1 \over 2} \{ 1+g(\alpha)+kf(\alpha)\cos{\phi}
  \nonumber\\
  & &\pm \sqrt{(1+g(\alpha)+kf(\alpha)\cos{\phi})^2-4g(\alpha) }\}.
\end{eqnarray}
Note that $\lambda^*_+$ becomes unity for $\phi=\pi/2$ or $3\pi
/2$ irrespective of $\alpha$ and $k$. In Fig.~\ref{fig:bifur}(b)
the logarithm of $|\lambda^*_\pm|$ is plotted as a function of
$\phi$ for $\alpha=0.3$ and $k=0.5$. We will use logarithms of
eigenvalues as a measure of the stability of a fixed point.
Stability along the eigendirection $\vece_+$ changes at $\phi=\pi
/2$ and $\phi=3\pi/2$. For $\phi<\pi/2$ or $\phi>3\pi/2$, the
fixed point $\vecd^*$ becomes a saddle point, at which trajectory
is attracted along $\vece_-$ eigendirection and repelled along
$\vece_+$ eigendirection, since $\ln|\lambda^*_+|>0$,
$\ln|\lambda^*_-|<0$ and $\lambda^*_\pm$ is real. For ${\pi/2} <
\phi < {3\pi/2}$ and $\ln|\lambda^*_+|<0$, the region of $\phi$ is
divided into two cases. First case is when logarithms of two
eigenvalues are negative and different. In this region, $\vecd^*$
become a stable node, which makes trajectories attracted along
$\vece_+$ to the fixed point, because $\ln|\lambda^*_-| <
\ln|\lambda^*_+| < 0$. For the other case, when
$\ln|\lambda^*_+|=\ln|\lambda^*_-| < 0$, two eigenvalues are
complex conjugated pair whose absolute values are less than 1.
Therefore, $\vecd^*$ becomes a stable spiral (stable foci) and
trajectory will be attracted to $\vecd^*$ rotating around the
fixed point.

Similarly for the other fixed point $\vecd^\circ$, the Jacobian
matrix is
\begin{eqnarray}
\label{eq:jacobian2}
      \jacobi^\circ(\phi)=\left(
        \begin{array}{cc}
        1 & f(\alpha) \\
        -k\cos\phi ~&~ g(\alpha)-kf(\alpha)\cos \phi
        \end{array}
        \right),
\end{eqnarray}
and its eigenvalues are
\begin{eqnarray}
  \label{eq:eigenvalue2}
  \lambda^\circ_\pm &=& {1 \over 2} \{ 1+g(\alpha)-kf(\alpha)\cos{\phi}
  \pm \nonumber\\
  & &~~~~\sqrt{(1+g(\alpha)-kf(\alpha)\cos{\phi})^2-4g(\alpha) }\}.
\end{eqnarray}
Same as the above case of $\vecd^*$, in Fig.~\ref{fig:bifur}(c)
$\ln|\lambda^\circ_+|$ is zero at $\phi=\pi/2$, $3\pi/2$ irrespective
of $\alpha$ and $k$. However, the situation of stability is
reversed. As shown in Fig.~\ref{fig:bifur}(b) and (c), $\vecd^*$
becomes stable spiral or node(saddle) and $\vecd^\circ$ becomes
saddle(stable spiral or node) at $\pi/2 <\phi <3\pi/2$~~($0<\phi
<\pi/2$ or $3\pi/2< \phi < 2\pi$). As a result
Fig.~\ref{fig:bifur}(a) shows corresponding bifurcation diagram
and $\vecd^*$ and $\vecd^\circ$ exchange the stability through
transcritical bifurcation at $\phi=\pi/2$, $3\pi/2$ respectively.
Note that $\vecd^*$ is constant but changes its stability with the
variation of $\phi$.

For some value of $\phi$ close to $\pi/2$ or $3\pi/2$, an attracting
closed orbit and a basin boundary between the stable fixed point and
the closed orbit were found in the phase portrait. Indeed, we have
obtained a homoclinic bifurcation which can be found in a
dissipative pendulum with constant torque\cite{guckenheimer}. In
our analysis, however, we have recognized that the existence of
basin boundary as a result of global bifurcation does not affect
our present results.

\subsection{Synchronization and On-off Intermittency}
\label{subsec:syncmanifold}
%%%%%%%%%%%%%%%%%%%%%%%%%%%%%%%%%%%%%%%%%%%%%%%%%%%%%%%%%%%%%%%
%
% Stability of synchronization manifold
%

Our main interest in this section is the mechanism of the
synchronization of two particles. Therefore, we will return to
Eq.~(\ref{eq:whole system}), where $\phi_n$ changes for every
iteration, and consider the stability of synchronized state based
on the study of previous section. The synchronized state corresponds
to $\vecd^*$ in $\vecf_{\phi_n}$. Because $\vecd^*=(0,0)$ is a
fixed point of $\vecf_{\phi_n}$, trajectory starting from
$\vecz_0=(\vecr_0,\vecd^*)$ belongs to a class of a particular
solution labelled by $(\vecr_0,\vecd^*)$ where $\vecr_0$ is
an arbitrary chosen vector. This class of trajectories constitute
2-dimensional invariant subset $\manm$ embedded in the whole
four-dimensional phase 
space. Because Eq.~(\ref{eq:whole system}) contains random
variable $\phi_n$, it will fill the whole hyperplane satisfying
$\vecd=\vecd^*$ and $\manm$ will be identical to the hyperplane.
We say a subspace is \emph{invariant} if a trajectory starting in
the subspace always remains in the same subspace.

Let us consider a situation where a trajectory, which is initially
located outside $\manm$, and $\vecz_n$ happens to be located in
the vicinity of $\manm$ for some $n$. Whether $\vecz_{n+1}$ will
be attracted or repelled from $\manm$ is determined by the
Jacobian matrix of subsystem $\vecf_{\phi_n}$ at $\vecd^*$ given
by Eq.~(\ref{eq:linear map}). The Eq.~(\ref{eq:linear map}) can be
mapped into a random walk when we take logarithm of absolute value
on each side of the equation as follows:
\begin{eqnarray}
  \label{eq:random walk}
  \ln|\vecd_{n+1}|=\ln\left|\jacobi^*(\phi_n)
    \cdot \hat{\vecd}_n \right|+\ln|\vecd_n|.
\end{eqnarray}
Note that we have divided $\vecd_n$ into $|\vecd_n|$ and unit
vector $\hat{\vecd}_n=(\cos\theta_n,\sin\theta_n)\transpose$,
where $\theta_n=\tan^{-1}(v_n/u_n)$. Then, one can regard
$\ln|\vecd_n|$ as $n$-th displacement of the random walk, and
$\ln|\jacobi^*(\phi_n)\cdot\hat{\vecd}_n|$ as its $n+1$-th step
width. Asymptotic behavior of the trajectory near $\manm$ is
determined by the average of the step width, i.e., the bias of
random walk. This bias is called transverse Lyapunov exponent
$h_\perp$, which measures global stability of $\manm$, defined as
\begin{eqnarray}
  \label{eq:transeveral liapunov}
  h_\perp=\int d\phi d\theta ~\rho(\phi,\theta)
  \ln\left| \jacobi^*(\phi) \cdot \hat{\vecd}(\theta)
  \right|,
\end{eqnarray}
where $\rho(\phi,\theta)$ is probability distribution of $\phi$ and
$\theta$.

If $h_\perp < 0$, i.e., the average of step width is negative, the
displacement of the random walk will decrease in time, and $|\vecd_n|$
will approach to zero in proportion to $e^{nh_\perp}$. Therefore
$\manm$ is transeversly stable, and the whole trajectories, which
initially start from outside of $\manm$, will be attracted to $\manm$
asymptotically. If $h_\perp > 0$, the small distance
between the trajectory and $\manm$ will increase exponentially and
at last will be affected by nonlinear term, which will reinject the
trajectory to the neighborhood of $\manm$. In this case $\manm$ is
called transeversly unstable. The transition from the former
case to latter one according to the variation of system paramter
has been investigated by others, and is called ``blowout''
bifurcation\cite{ott:pla:94,ashwin:pla:94,lai:pre:95}.

\begin{figure}[htbp]
  \centering
  \epsfig{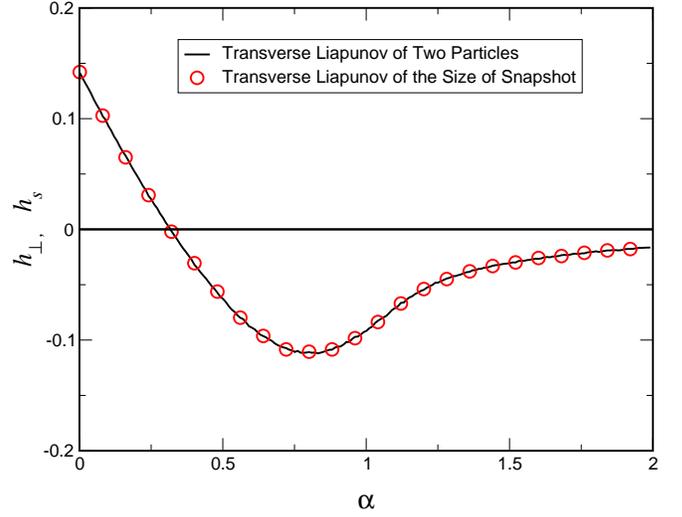}
  \caption{$h_\perp$(solid line) and ~$h_s$(circle) vs $\alpha$ when k=0.5.}
  \label{fig:lyapunov}
\end{figure}

Practically we can calculate $h_\perp$ as follows:
\begin{eqnarray}
  \label{eq:lyapunov}
  e^{nh_\perp}&=&\lim_{n\rightarrow\infty}{\sqrt{u_n^2+v_n^2} \over
    \sqrt{u_0^2+v_0^2}} \nonumber\\
  &=&\lim_{n\rightarrow\infty}\left|
    {\mathbf{J}}^*(\phi_n){\mathbf{J}}^*(\phi_{n-1})
    \cdots {\mathbf{J}}^*(\phi_0)
    \cdot \hat{\vecd}_0\right|\nonumber\\
  \rightarrow h_\perp &=&\lim_{n\rightarrow\infty} {1\over n} \sum_i
    \ln \left|{\mathbf{J}}^*(\phi_i)\cdot \hat{\vecd}_i\right|.
\end{eqnarray}
$h_\perp$ vs $\alpha$ graph is shown in the Fig.~\ref{fig:lyapunov} which
coincides with the previous work\cite{yoc:prl,yoc:physica}.

\begin{figure}[htbp]
    \centering
    \epsfig{file=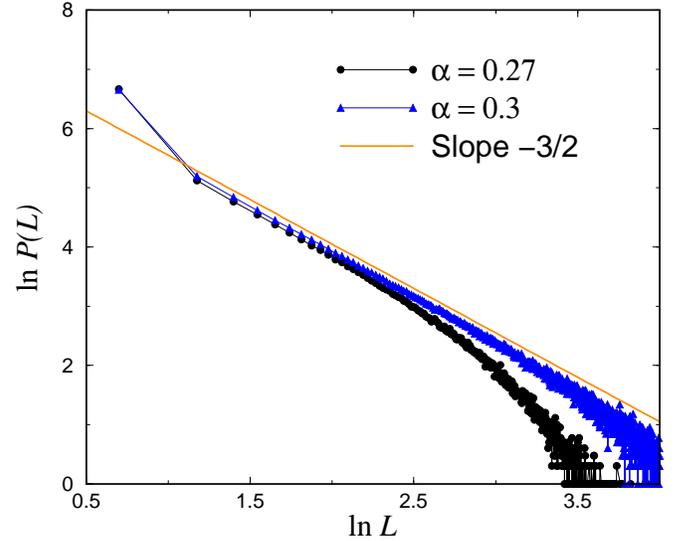,width=\figw}
    \caption{Laminar distribution for two particles.
    Horizontal axis is length
    of laminar in logarithmic scale. Vertical axis is logarithm
    of probability
    of laminar length in arbitrary units.
    Filled circle means probability for
    $\alpha=0.27$ and filled triangle for $\alpha=0.3$.
    Gray straight line
    have $-3/2$ slope. The case of $\alpha=0.27$ shows
    exponential shoulder which
    is typically shown in on-off intermittency models.}
    \label{fig:laminar}
\end{figure}

It is also known that when $h_\perp$ is slightly positive, some
dynamical variables of the system can exhibit an extreme type of
temporal intermittent bursting behavior: on-off
intermittency\cite{heagy:pre:94,platt:prl:94}. The essential
ingredients of on-off intermittency are : (a) a hyperplane
should contain invariant subset and (b) trajectory near hyperplane
shows additive random walk in log domain depending on
the local stability of hyperplane along the transverse direction
of $\manm$\cite{ding:pre:95,platt:prl:94}. Trajectory in large negative
value in log domain will be shown approximately a constant in real
scale, and corresponds to the `off' state. In other case positive
or small negative value in log domain corresponds to `on' state.
In the pairs of particles, we have shown that $\manm$ is the
hyperplane containig invariant subset, and $\ln|\vecd_n|$ exhibits
random walk  near $\manm$ according to the the stability of
$\jacobi^*$. Therefore we can conclude that there exists on-off
intermittency. On the other hand,  according to a result based on
the study of a random walk, it is known that the probability of
laminar length($L$) is proportional to $L^{-3/2}e^{-L/L_s}$, where
$L_s$ is the length at which the systematic drift due to the bias
and the diffusion spread of random variable without bias become
comparable\cite{yang:pre:96,ding:pre:95,heagy:pre:94}.
Fig.~\ref{fig:laminar} shows characteristic scaling of on-off
intermittency which $P(L)$ scales as $L^{-{3 \over 2}}$.

In this section, we have shown the mechanism in the
synchronization of two particles using skew-product structure and
stability analysis of $\manm$ and verified structure of on-off
intermittency.

%%%%%%%%%%%%%%%%%%%%%%%%%%%%%%%%%%%%%%%%%%%%%%%%%%%%%%%%%%%%%%%%%%%
%%%%%%%%%%%%%%%%%%%%%%%%%%%%%%%%%%%%%%%%%%%%%%%%%%%%%%%%%%%%%%%%%%%
%%%%%%%%%%%%%%%%%%%%%%%%%%%%%%%%%%%%%%%%%%%%%%%%%%%%%%%%%%%%%%%%%%%
\section{Synchronization of Ensemble of $N$ Particles}
\label{sec:randomwalk}

\begin{figure}[htbp]
    \centering
    \epsfig{file=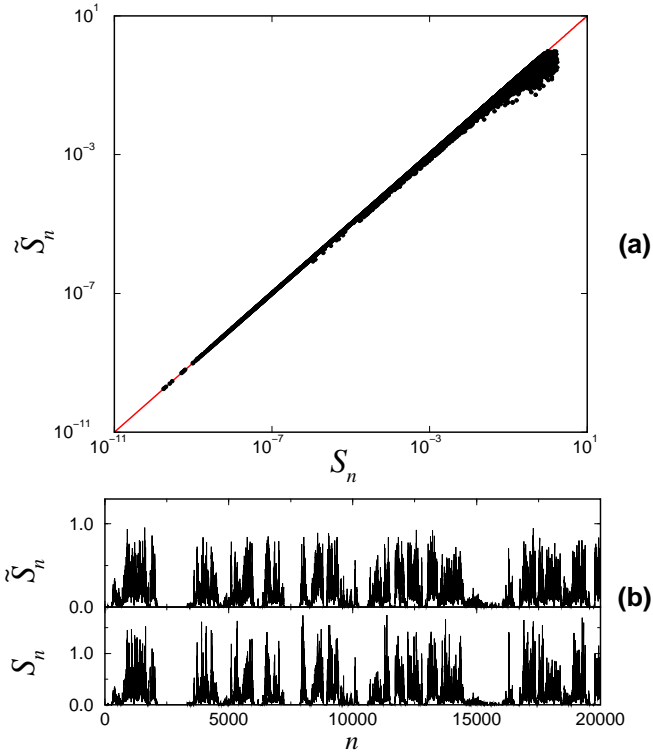,width=\figw}
    \caption{(a) Horizontal and vertical axis are $S_n$ and
    $\tilde{S}_n$ in logarithmic scale, respectively. Both
    axes are logarithmic scale. Solid
    dots are data from 1000 particles when $\alpha=0.3$ and
    grayed line is diagonal line. Data close
    to zero lies on the diagonal line and
    that means $S_n$ and $\tilde{S}_n$ become identical.
    (b) shows time series of $S_n$ and
    $\tilde{S}_n$ for same period. Although amplitudes are slightly
    different each other, timings of bursting
    and laminar period coincide.}
    \label{fig:size of snapshot}
\end{figure}

In this section, we will study the behavior of ensemble of $N$
particles. As described in previous section, on-off intermittency
can be characterized by a biased random walk. To establish the
link with a random walk with the size evolution of the snapshot attrctor, let
us consider the relation between the ED and the size of the
snapshot attractor. YOC considered the size of the snapshot
attractor $S_n$ as the \textit{r.m.s.} (root mean square) value of distance
from the center of mass of ensemble as follows:
\begin{eqnarray}
  \label{eq:attractorsize_old}
  S_n = \sqrt{{1\over N}\sum_{i=1}^N {(y^{i}_n - \bar{y}_n)^2 }} ,
\end{eqnarray}
where $y^{i}_n$ are coordinates of $i$-th
particle($i=1,2,\ldots,N$) and $\bar{y}_n=N^{-1}\sum_i
y^{i}_n$. However, $\bar{y}_n$ is not so good choice to establish
the random walk relation between $S_n$ and $S_{n-1}$. Since
$\bar{y}_n$ is defined as average of $y^{i}_n$ and $S^2_n$ can be
expressed by $\bar{y}_n$ and $\bar{(y_n)^2}$, it is difficult to
obtain $S_{n-1}$ from the definition of $S_n$ in the limit of
$S_n\rightarrow 0$.
In order
to use the result from the previous section, we
consider the modified measure $\tilde{S}_n$ which is the average
length not from the center of the ensemble $\bar{y}_n$
but from an arbitrarily chosen reference particle position $y^r_n$
in the ensemble. Our modified measure is defined as
\begin{eqnarray}
  \label{eq:attractorsize}
  \tilde{S}_n = \sqrt{{1 \over N} \sum_i {(y^{i}_n-y^r_n)^2}}=
  \sqrt{{1 \over N} \sum_i {(v^{i}_n)^2}},
\end{eqnarray}
which is \textit{r.m.s.} distance in $y$ direction, where
$v^{i}_n=y^{i}_n-y^r_n$. The behavior of $S_n$ and $\tilde{S}_n$
can be seen in Fig.~\ref{fig:size of snapshot} for the case
$N=1000$, $\alpha=0.3$, and $k=0.5$. Fig.~\ref{fig:size of
snapshot}(b) shows the amplitude are slightly different, but
bursting and laminar behaviors coincide. Moreover, as shown in
Fig.~\ref{fig:size of snapshot}(a), as $S_n$ approaches to 0,
$\tilde{S}_n$ becomes identical to $S_n$. Therefore $\tilde{S}_n$
may be replaced with $S_n$ when one investigates the size of the
snapshot attractor near synchronization transition point. Since our measure $\tilde{S}_n$ can
be decomposed into \textit{r.m.s.} value of the ED variable $v^{i}_n$, the
relation between $\tilde{S}_{n+1}$ and $\tilde{S}_n$ can be
obtained from the relation between $v_{n+1}$ and $v_n$. In the
measure of $\tilde{S}_n$, the existence of the hyperplane
containing invariant set is guaranteed by invariant hyperplane of two
particles system, that is the hyperplane stands for the state when all the
particles move synchronously.

\subsection{Uncoupled map lattice with homogeneous background}
\label{sec:yang-ding model}

We analyzed one dimensional simple case of globally coupled
logistic maps which was studied by
YD\cite{yangding:pre:94,snapshot:other}. Let us consider the
system of $N$ particles. They are located at
$y^{1}_n,\ldots,y^{N}_n$, and their dynamics are described by
\begin{eqnarray}
  \label{eq:yangdingmodel}
  y^{i}_{n+1}=z_n y^{i}_n ( 1-y^{i}_n ),
\end{eqnarray}
where $z_n = a\xi_n+b$ and $\xi_n$ is a random variable uniformly
distributed in the interval $(0,1]$. Note that this system is
identical to the model studied by HPH\cite{heagy:pre:94} to
explain on-off intermittency. When $a=1$ is fixed and $b$ is
varied, they reported that the transition from non-synchronous state
to synchronous one is found at $b_c=2.82$, and when $b>b_c$, the
size evolution of the snapshot attractor exhibit on-off intermittency. We will
explain this transition behavior with our measure $\tilde{S}_n$
and the ED. If we consider the difference $v^i_n$ between
$y^i_n$ and $y^r_n$ such that
\begin{eqnarray}
  \label{eq:yangding-ED}
  \vn1i &=& y^{i}_{n+1}-y^r_{n+1} \nonumber\\
  &=& z_n \vni ( 1-2y^r_n-\vni),
\end{eqnarray}
then,
\begin{eqnarray}
  \label{eq:snapshot yangding}
  \tilde{S}^2_{n+1} &=& {1\over N}\sum^N_{i=1}(\vn1i)^2 \nonumber\\
  &=& {1\over N}\sum^N_{i=1}z^2_n(\vni)^2(1-2y^r_n-\vni)^2.
\end{eqnarray}
If we consider the limit $\vni\rightarrow 0$ and neglect higher order
terms $\sum_i (\vni)^3$ and $\sum_i (\vni)^4$ near the transition region,
Eq.~(\ref{eq:snapshot yangding}) becomes
$\tilde{S}_{n+1} \simeq |z_n(1-2y^r_n)|\tilde{S}_n$. Then, we have
obtained the following relation by taking logarithm on both side.
\begin{eqnarray}
  \label{eq:randomwalk yangding}
  \ln \tilde{S}_{n+1} = \ln|z_n(1-2y^r_n)|+\ln \tilde{S}_n.
\end{eqnarray}
This relation can be interpreted such that the size evolution of the
snapshot attractor in log domain is governed by the random walk
with a step size $\ln|z_n(1-2y^r_n)|$. Similarly to
Eq.~(\ref{eq:transeveral liapunov}) the transverse Lyapunov exponent is
\begin{eqnarray}
  \label{eq:lambta_t yangding}
  h_\perp = \int dy~\rho(z,y) \ln|z(1-2y)|,
\end{eqnarray}
where $\rho(z,y)$ is invariant probability density. This leads to the
same transition point which is calculated by YD.
One should notice that the random variable is
multiplied \emph{commonly} to every particles in the summation of
Eq.~(\ref{eq:snapshot yangding}), and this leads to 
on-off intermittency of $\tilde{S}_n$.

\subsection{RZ map}
\label{sec:snapshot of RZ map}

\begin{figure}[htbp]
  \begin{center}
    \epsfig{file=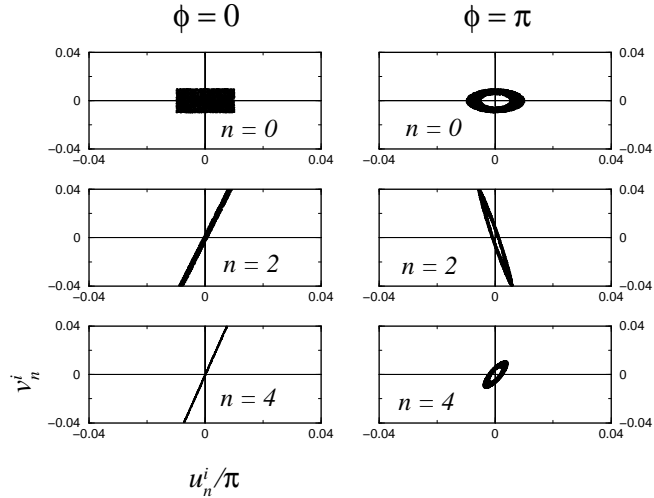,width=\figw}
    \caption{Evolution of ensemble of 5000 particles for constant
      $\phi$ when $\alpha=0.3$ and $k=0.5$. Left column of
      three graphs are initial evolution of ensemble with $\phi=0$,
      where fixed point at $\vecd^*$ is saddle. We start with random initial
      points distributed within $-0.01<u_0/\pi<0.01$ and
      $-0.01<v_0<0.01$ uniformly in order to show the behavior near
      saddle. From upper to lower graphs, shape of ensemble stretch
      along unstable manifold as time flows. Contrast to left ones,
      right column are that of ensemble when $\phi=\pi$ and fixed point
      is spiral. The distribution, which is initially located
      in a ring, rotate around origin with
      contracting its area. Note that in these graphs coordinates are
      $u^i_n$ and $v^i_n/\pi$, i.e. relative displacement from the
      reference particle.}
    \label{fig:saddle-spiral evolve}
  \end{center}
\end{figure}

In the case of the RZ map, we consider the case of $\vni\simeq 0$.
From Eq.~(\ref{eq:error-dynamics}),
\begin{eqnarray}
    \label{rz-snapshot-approx}
    \vn1i &=& g(\alpha)\vni+2k\cos\left(\phi_n+ \un1i/2
       \right)\sin\left(\un1i/2\right)\nonumber\\
    &=& g(\alpha)\vni+k\cos{\phi_n}
    \sin[(\tni+f(\alpha))\vni]\nonumber\\
    & &-2k\sin{\phi_n}\sin^2\left[(\tni +f(\alpha))
      \vni/2\right]\nonumber\\
    &\simeq&
    [g(\alpha)+k\cos\phi_n(\tni+f(\alpha))]\vni+O((\vni)^2),\nonumber\\
\end{eqnarray}
where $\tni=\uni/\vni$ and $\un1i=[\tni+f(\alpha)]\vni$.
Therefore,
\begin{eqnarray}
    \label{eq:rz-snapshot}
    \tilde{S}^2_{n+1}&=&{1\over N}\sum_i [\vn1i]^2\nonumber\\
    &\simeq& {1\over N}\sum_i
    \{g(\alpha)+k\cos\phi_n[\tni+f(\alpha)]\}^2(\vni)^2\nonumber\\
    & &+{1\over N}\sum_i O[(\vni)^3].
\end{eqnarray}
Then, Eq.~(\ref{eq:rz-snapshot}) have different result comparing
with previous case, because the summation in
Eq.~(\ref{eq:rz-snapshot}) does not have a common driving. As a
result, it seems that one could not deduce $\tilde{S}_n$ from the
right side. However, we have recognized the crucial property that
the value of $\tni$ for all $i$ becomes identical when
$\vni\simeq 0$, which will be explained in detail in the following. That
properties means that the whole particles aligned in a 
line passing through reference particle, because $\tan\theta^i_n$
is the slope of a line connecting position of the reference
particle to that of the $i$-th particle. In Sec.~\ref{sec:sync},
we have shown that $\vecd^*$ becomes a saddle or a spiral for most
$\phi$. When $\vecd^*$ is saddle, particles are repelled along
unstable direction. However, since
$|\ln|\lambda^*_+||<|\ln|\lambda^*_-||$, particles are attracted
fast along $\vecd_-$ and repelled slowly along $\vecd_+$.
Therefore, particles are scattered around along unstable
eigendirection after a few iteration, and become a narrow stripe
along a straight line from the reference particle. When $\vecd^*$
is spiral, we have shown that the particles around $\vecd^*$
rotate around the fixed point without changing the shape of
distribution. Fig.~\ref{fig:saddle-spiral evolve} shows the
behavior of particles near $\vecd^*$ for constant $\phi$ when
$\vecd^*$ is saddle or spiral. After particles are scattered along
a line, the Jacobian matrix maps the line passing through the
position of reference particle to another line passing through
that of reference particle. Therefore, we could approximate
$\cot\theta_n=\cot\theta^{1}_n=\cot\theta^{2}_n=\ldots=\cot\theta^{N}_n$
for $\tilde{S}_n\simeq 0$.

\begin{figure}[htbp]
  \begin{center}
    \epsfig{file=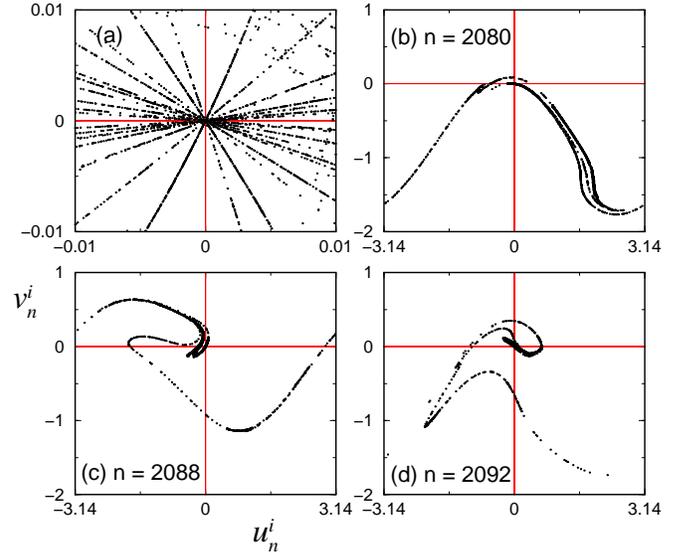,width=\figw}
    \caption{Snapshot of 5000 particles when $\alpha=0.3$ and
      $k=0.5$. Initially, particles are distributed in the region
      $-0.5<u_0<0.5$ and $-0.5<v_0<0.5$. (b),(c) and (d) shows a fractal
      distribution of particles when $n=2080,2088$ and $2092$
      respectively. (a) shows small area of snapshot attractor for
      $n=2080,2081,\ldots,2099$ in a same graph. In this region particles
      are aligned in a line for each instant time when particle are close
      to origin. Each line stands for a snapshot for a given instant
      time. Note that in these graphs coordinates are $u^i_n$ and
      $v^i_n$, i.e. relative displacement from the reference particle.}
    \label{fig:tangent theta}
  \end{center}
\end{figure}

\begin{figure}[htbp]
  \begin{center}
    \epsfig{file=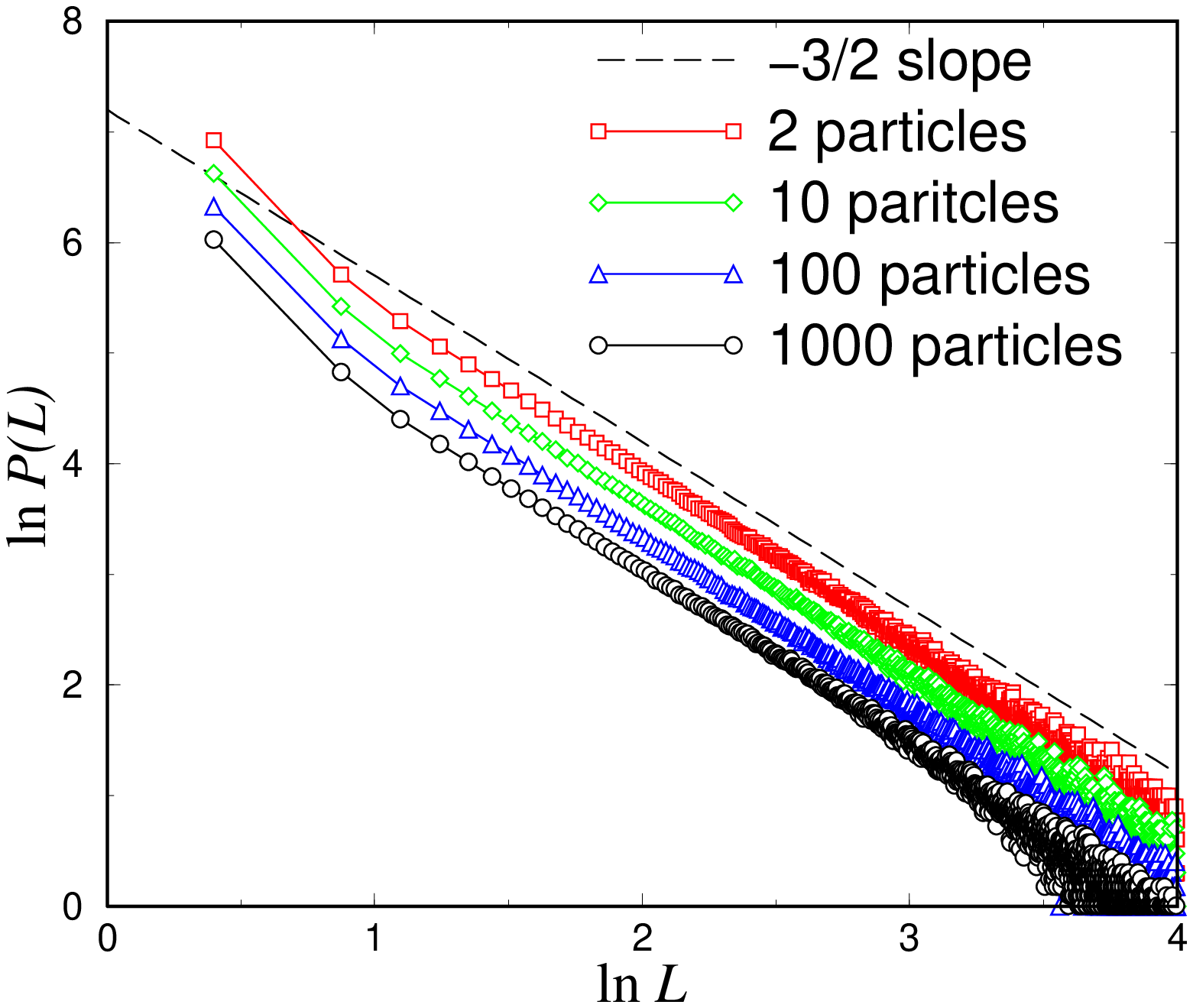,width=\figw}
    \caption{Probability distribution $P(L)$ of laminar length
      $L$ when $\alpha=0.3$ and $k=0.5$. We choose the threshold of on
      event as $10^{-4}$. The unit of probability is arbitrary.}
    \label{fig:laminar-ensemble}
  \end{center}
\end{figure}

This result is consistent with the
finding of YOC that Lyapunov dimension becomes unity on chaotic
side of transition. This represents that ensemble of $N$ particles
is embedded in one-dimensional manifold. Also from an evolution of
uniformly distributed ensemble, we have found that near the fixed
point all the particles are aligned in a line from the position of
reference particle as shown in Fig.~\ref{fig:tangent theta}.
Therefore, we can write as follow:
\begin{eqnarray}
  \label{eq:snapshot onoff}
  \tilde{S}_{n+1}=\left|g(\alpha)+
  k\cos\phi_n(\cot\theta_n+f(\alpha))\right|\tilde{S}_n.
\end{eqnarray}
By taking logarithm on both side, we find
\begin{eqnarray}
  \label{eq:log snapshot onoff}
  \ln&\tilde{S}_{n+1}&= \nonumber \\
  & &\ln\left|g(\alpha)+
  k\cos\phi_n(\cot\theta_n+f(\alpha))\right|+\ln\tilde{S}_n.
\end{eqnarray}
As a result, Eq.~(\ref{eq:log snapshot onoff}) shows a random walk
process in log domain. The transition point of the size of the
snapshot attractor is determined by transverse Lyapunov exponent $h_s$, which
is the average step size in 
Eq.~(\ref{eq:log snapshot onoff}) as follows:
\begin{eqnarray}
  \label{eq:liapunov snapshot}
  h_s=\lim_{n\rightarrow\infty}\frac{1}{n}\sum_{i=1}^n
  \ln\left|g(\alpha)+  k\cos\phi_n(\cot\theta_n+f(\alpha))\right|.
\end{eqnarray}
In Fig. \ref{fig:lyapunov}, the $h_s$, which is numerically obtained
by using the original map (\ref{eq:yoc map}), shows the same result of
$h_\perp$ which is calculated in case of two particles.
In this case invariant hyperplane is a subspace satisfying $\vni=0$
and $\uni=0$ for all $i$. By the hyperplane and the previous random
walk relation, existence on-off intermittency of $\tilde{S}_n$ is
verified.
Furthermore, in the plot of laminar distribution of $\tilde{S}_n$,
the distribution is not affected by the number of particles and
shows same scaling of the two particles case,
as depicted in Fig.~\ref{fig:laminar-ensemble}, which shows the
consistency of our results.

%%%%%%%%%%%%%%%%%%%%%%%%%%%%%%%%%%%%%%%%%%%%%%%%%%%%%%%%%%%%%%%%%%%%%%%%%
%%%%%%%%%%%%%%%%%%%%%%%%%%%%%%%%%%%%%%%%%%%%%%%%%%%%%%%%%%%%%%%%%%%%%%%%%
\section{Summary and Discussion}
\label{sec:summary & discussion}

In this paper, we have considered the mechanism of synchronization
in random dynamical system. As a specific example, we have investigated
the random Zaslavsky map in detail. From the ED of two particles, we
have studied the bifurcation structure of invariant manifold
and found transcirtical bifurcation between saddle and stable
node. Our results are consistent with those of YOC, but are more
explicit in the perspective of on-off intermittency structure. In
case of ensemble of many particles in the random Zaslavsky map, the
long standing problem to map the size evoultion of the snapshot attractor
into a random walk is at last resolved by exploiting our modified
definition of the size of the snapshot attractor, which is slightly different
from the previously proposed one. From this success of mapping into
a random walk, we have obtained qualitative understanding why this
manifestation of intermittency has the same critical exponent, which is
found from numerical simulation as that of the typical on-off
intermittency. We emphasize that this mapping becomes possible
when the ED variable are distributed on a line along the unstable
direction of saddle. For further investigation, it will be
interesting to study the other systems, that seem
to be difficult to map into the form of a random walk.

%%%%%%%%%%%%%%%%%%%%%%%%%%%%%%%%%%%%%%%%%%%%%%%%%%%%%%%%%%%%%%%%%%%%%%
%%%%%%%%%%%%%%%%%%%%%%%%%%%%%%%%%%%%%%%%%%%%%%%%%%%%%%%%%%%%%%%%%%%%%%
%% Acknolwegement
\acknowledgements We thank P. Kang for valuable
discussion. This work was supported by Creative Research
Initiatives of the Korean Ministry of Science and Technology. Two
of us (I. K. \& Y.-J. P.) acknowledge support from the Ministry of
Education, BK21 Project No. D-1099.

%%%%%%%%%%%%%%%%%%%%%%%%%%%%%%%%%%%%%%%%%%%%%%%%%%%%%%%%%%%%%%%%%%%%%%%%%%%%%%%%%%%%
%%%%%%%%%%%%%%%%%%%%%%%%%%%%%%%%%%%%%%%%%%%%%%%%%%%%%%%%%%%%%%%%%%%%%%%%%%%%%%%%%%%%

\begin{thebibliography}{12}
\bibitem{ed-fp} Fixed points can be found analytically.
  When
  $v_nf(\alpha)=2l\pi$ for $l=\pm 1,\pm 
  2,\ldots$, after applying condition
  $u_{n+1}=u_n$ and $v_{n+1}=v_n$, we obtain $\sin u_n = 2l\pi\alpha/k +
  \sin\phi$ from Eq~(\ref{eq:error-dynamics}). For given $l$ this
  equation has solutions when 
  $\alpha<1/4\pi l$. When $l=\pm 1$, soultion exists for
  $\alpha<1/4\pi\simeq 0.16$.
  However these fixed points does not affect our
  analysis because $0.16$ is fairly apart from the transition point $\alpha_c$.
  When $l=\pm 2,\pm 3, \ldots$, the range of $\alpha$ where solutions exist
  become smaller than the range in the case of $l=\pm 1$, therefore
  all the other fixed points except the case of $v_nf(\alpha)=0$ are
  irrelevant to our analysis.
\end{thebibliography}
\end{document}